\documentclass[12pt]{article}

\usepackage{graphicx,float}
\usepackage{epsf,amsmath,bbold,amsfonts}
\usepackage{appendix}
\usepackage{amssymb}

\usepackage{paper2e}
\usepackage{xcolor}
\usepackage{hyperref}
\hypersetup{colorlinks=false,pdfborderstyle={/S/U/W 1}}

\def\a{\alpha}
\def\b{\beta}
\def\c{\chi}

\def\d{\delta}
\def\D{\Delta}

\def\eps{\varepsilon}
\def\f{\frac}
\def\g{\gamma}

\def\G{\Gamma}

\def\l{\left}

\def\la{\langle}
\def\ra{\rangle}

\def\mc{\mathcal}

\def\m{\mu}
\def\n{\nu}

\def\nn{\nonumber}

\def\p{\partial}

\def\r{\right}

\def\z{\zeta}

\def\be{\begin{equation}}
\def\ee{\end{equation}}

\def\bea{\begin{eqnarray}}
\def\eea{\end{eqnarray}}

\def\ba{\begin{array}}
\def\ea{\end{array}}

\def\bc{\begin{center}}
\def\ec{\end{center}}

\def\bl{\begin{flushleft}}
\def\el{\end{flushleft}}

\def\br{\begin{flushright}}
\def\er{\end{flushright}}

\def\bi{\begin{itemize}}
\def\ei{\end{itemize}}

\def\bt{\begin{tabular}}
\def\et{\end{tabular}}

\begin{document}

\begin{titlepage}
\vspace{5cm}

\vspace{2cm}

\begin{center}
\bf \Large{ Spontaneously Broken Conformal Symmetry: \\
Dealing with the Trace Anomaly}

\end{center}

\begin{center}
{\textsc {Roberta Armillis, Alexander Monin, Mikhail Shaposhnikov}}
\end{center}

\begin{center}
{\it Institut de Th\'eorie des Ph\'enom\`enes Physiques, \'Ecole Polytechnique F\'ed\'erale de Lausanne, CH-1015, Lausanne, Switzerland}
\end{center}

\begin{center}
\texttt{\small roberta.armillis@epfl.ch} \\
\texttt{\small alexander.monin@epfl.ch} \\
\texttt{\small mikhail.shaposhnikov@epfl.ch} \\
\end{center}

\vspace{2cm}

\begin{abstract}

The majority of renormalizable field theories possessing the scale invariance at the classical level exhibits the trace anomaly once quantum corrections are taken into account. This leads to the breaking of scale and conformal invariance. At the same time any realistic theory must contain gravity and is thus non-renormalizable. We show that discarding the renormalizability it is possible to construct viable models allowing to preserve the scale invariance at the quantum level. We present explicit one-loop computations for two toy models to demonstrate the main idea of the approach. Constructing the renormalized energy momentum tensor we show that it is traceless, meaning that the conformal invariance is also preserved.

\end{abstract}

\end{titlepage}

\newpage
\section{Introduction}

Symmetry is one of the most important notions in physics. The majority of all theories is based on this concept and at the same time it is one of the most powerful   guiding principles. Certain features of a system  such as  conservation laws or Ward identities  appear as natural  consequences of symmetries only. Conformal invariance is an example of a symmetry restricting drastically the form of the observables (correlators), allowing in many cases to make long reaching conclusions.

Scale and conformal invariant field theories (SFT and CFT) arise in different parts of physics. These theories are very specific, meaning that conformal symmetry is present in the system only for certain values of parameters. In solid state physics CFTs are used to describe systems at the critical temperature \cite{Kadanoff:1966wm,Polyakov:1970xd,Belavin:1984vu}, while in high energy physics they appear as UV and/or IR limits (fixed points)~\cite{Komargodski:2011xv,Komargodski:2011vj,Luty:2012ww,Fortin:2012hc,Fortin:2012hn} or in supersymmetric field theories such as $\mc{N}=4$ SYM~\cite{Sohnius:1981sn,Aharony:1999ti}.
CFTs have been  extensively studied since long ago~\cite{Wess:1960,Mack:1969rr,Ferrara:1973eg}. For a more systematic and  pedagogical review we refer an interested reader to~\cite{DiFrancesco:1997nk,Blumenhagen:2009zz}.

Investigating CFTs it is usually assumed that not only the Lagrangian but also the vacuum is invariant under conformal transformations. In the case of non-Lagrangian formulations of CFTs (the bootstrap approach) the symmetry is made manifest through the appropriate action of the conformal group on the correlation functions, i.e. Ward identities. On the other hand phenomenologically viable  theory for particle physics has to exhibit the conformal symmetry breaking in one or another way to properly describe the real world. For instance this symmetry can be  explicitly broken by mass terms or via dimensional transmutation, as in QCD, then it is exact only in the high energy limit. The way to reconcile two seemingly contradicting conditions, namely, the existence of exact scale invariance at the Lagrangian level and the generation of a scale, is to consider the spontaneous breaking of the symmetry. In this case a vacuum expectation value (vev) of some operator provides the necessary dimensionful parameter.

%Two seemingly contradicting statements of scale invariance and difference between IR/UV theories are easily reconciled if the symmetry is spontaneously broken:

In~\cite{Shaposhnikov:2008xi,Shaposhnikov:2008xb} the advantages of considering scale invariant theories with spontaneous symmetry breaking were presented. In these theories the presence of yet another shift symmetry makes the Higgs mass stable against radiative corrections and the problem of cosmological constant acquires a different meaning. Therefore, it is desirable to have a theory which exhibits scale invariance and at the same time incorporates the Standard Model (SM) as its low energy limit.

There are two ways to approach the problem at hand. One is top-down, when starting from a UV complete theory with spontaneously broken scale invariance, one chooses its parameters in such a way that it reproduces the SM in the IR. However, we are not aware of any theory with spontaneously broken scale symmetry which would at least vaguely resemble the~SM. Therefore, we adopt another approach, namely, bottom-up. It means that, starting from the SM itself, it is modified in such a fashion as to make it scale invariant.

The way to do that at the classical level is rather obvious. Since the only dimensionful parameter in the model is the vev of the Higgs field, it is enough to consider it as deriving from the vev of a dynamical field $ X $, called dilaton~\cite{Coleman_Aspects}. More generally the procedure can be formulated as follows. It is known that a symmetry of the classical Lagrangian can be extended if one considers its nonlinear realization \cite{Coleman:1969sm,Callan:1969sn,Salam:1969rq,Salam:1970qk}. It means that one can compensate for the non-invariance of the Lagrangian by introducing the corresponding Goldstone modes, one for each broken generator, i.e. treating the symmetry as if it were spontaneously broken.

In the case of broken scale invariance there is only one broken generator. The corresponding Goldstone mode is the dilaton. If one considers the full conformal group which is broken down to the Poincar\'e group there are 5 broken generators corresponding to dilatations and special conformal transformations. However, in the case of space-time symmetries the naive counting of the Goldstone modes does not work \cite{Low:2001bw} in the same way as for global symmetries. In fact it is enough to introduce only one Goldstone boson corresponding to the scale symmetry~\cite{Isham:1970gz,Rattazzi:2003ea,Sundrum:1998sj} in order to have a conformally invariant system.

In doing so one faces a rather general problem. It is not uncommon that a symmetry is lost once quantum corrections are taken into account, a case generally referred to as the emergence of quantum anomalies.  Therefore, what seemed to be a symmetry of the classical action is not for the quantum system anymore. Usually it is said that the symmetry is not consistent with the regularization/renormalization procedure, meaning that, although the symmetry is apparent in the classical Lagrangian, it gets violated once the regularization is performed.

According to what is commonly believed about scale and conformal invariance, in most known {\it renormalizable} theories these symmetries are explicitly broken \cite{Coleman_Aspects}.  Therefore, pursuing the search of a  realistic conformally invariant theory at the quantum level may seem a hopeless task. However,  when dealing with non-renormalizable theories this question should be carefully reanalyzed. The renormalizability of the theory is not a necessary requirement in order for a theory to be realistic. It is enough to think that aiming at the construction of a complete theory lets gravity come into play, which is non-renormalizable. It is perfectly consistent to study non-renormalizable CFTs exhibiting spontaneous breaking of conformal invariance. Hence, one may look for an appropriate formalism preserving manifestly the symmetry order by order in perturbation theory, abandoning the renormalizability condition.

%Scale invariance is not an exception \cite{Coleman_Aspects}. The divergence of the corresponding current is not zero at the quantum level, but rather proportional to the $\beta$-function of  the theory at hand~\cite{Coleman_Aspects,Callan:1970ze,Brown:1979pq}\footnote{That being true in a case without additional flavor symmetry. The systems with additional flavor symmetry are thoroughly investigated in \cite{Luty:2012ww,Fortin:2012hn}.}.

In~\cite{Shaposhnikov:2008xi,Shaposhnikov:2008xb} it was shown how to construct a wide class of scale invariant at the quantum level non-supersymmetric {\it effective} field theories, which can describe particles and be predictive up to high energy scale \footnote{We will describe the method in detail in the next section.}. The approach is reminiscent of the one used in earlier works \cite{Englert:1976ep} and \cite{Wetterich:1987fm,Wetterich:1987fk}. The main idea is to choose a suitable regularization such that the symmetry in question is preserved.
%Naively, it seems that the dimensional regularization is a good candidate to achieve this. However, the parameters of the Lagrangian become dimensionful once extended beyond 4 space-time dimensions. 

%The authors of \cite{Shaposhnikov:2008xi,Shaposhnikov:2008xb} used the so called scale invariant (SI) prescription which makes the regularized Lagrangian manifestly scale invariant in an arbitrary number of dimensions by means of a dilaton. 
For all standard regularizations a mass scale appears. This scale can be identified with the renormalization scale $\mu$, with the cutoff $\Lambda$ or with the mass $M$ of the Pauli-Villars regulator. In one way or another a dimensionful quantity is responsible for the explicit breaking of the scale invariance. In order to advance in developing the formalism which keeps the theory scale invariant at each step, it is enough to use a scheme in which this generic mass parameter is replaced by a dynamical field. This renders the \emph{regularized} Lagrangian scale invariant by definition. In what follows this particular choice of prescription is called Scale Invariant (SI).

One important remark is in order. It may seem a trivial statement saying that the symmetry of the quantum system is preserved if the {\it regularization} respecting the symmetry is used.  However, regularizing a theory is not the end of the story. One has to renormalize it as well by adding appropriate counterterms. Generally speaking these terms may break the symmetry. A well known example is the Weyl anomaly~\cite{Capper:1974ic,Duff:1993wm} which in dimensional regularization originates from the non-invariance of the counterterms. 

In \cite{Shaposhnikov:2008xi} it was shown that the counterterms needed to remove the divergences at one loop are scale invariant. Therefore, the one-loop effective action is also scale invariant. Moreover, the regularized Lagrangian is also invariant under the special conformal transformations with parameter $a_\m$, acting on a generic spinless field $\Phi$ of scaling dimension $\Delta$ in the following way
%{\bf Therefore, the question of the invariance of the Lagrangian at the quantum level is now extended from the scale symmetry to the conformal symmetry. It means that we want to follow the same steps as for the scale invariance, but in this case dealing with the conformal invariance. Our starting point will be a generic Lagrangian, which will be regularized with the use of the SI prescription and then renormalized.} 
%, several questions remained unanswered and this is why we want to further investigate the consequences of using the SI prescription. 
%For example, the construction of the corresponding current (in this case related to the energy-momentum tensor) was not performed. It is also interesting to find out what happens to the conformal symmetry, apparently present at the classical level. Is it preserved at the quantum level?
\be
\Phi ' (x) = \f {1} {(1- 2 (a x) + a ^ 2 x ^ 2)^ {\D}} \Phi \l ( \f {x - a x^2} {1- 2 (a x) + a ^ 2 x ^ 2} \r ).
\label{conftrans}
\ee

One might wonder whether this symmetry is preserved at the quantum level as well. Our conjecture is that in flat space-time it is possible within the SI scheme -- which makes the {\it regularized} Lagrangian explicitly invariant -- to choose the symmetry preserving counterterms, therefore, rendering the quantum system conformal. This provides a way of building a model with spontaneously broken conformal invariance with a given low energy limit described by the classical Lagrangian. In the following sections we are going to show in detail that at one loop this is indeed the case.

For our purposes we are going to focus only on dimensional regularization (for a discussion of scale-invariant lattice realization
see \cite{Shaposhnikov:2008ar}), in which case it amounts to say that the {\it regularized} Lagrangian is invariant under the conformal transformations in an arbitrary number of dimensions $n$ which is achieved by multiplying the terms with dimensionful couplings by an appropriate power $\g$ of the dilaton field $X $.
%In (\ref{conftrans}), as usual,  $\Phi$ stands for a generic spinless field of scaling dimension $\Delta$, and the components of the four-vector $a$ are the parameters of the special conformal transformation.
We stress once more that our analysis is carried out in flat space time, leaving the study of systems coupled to gravity for future investigations.

%Our aim is to show here explicitly by means of a perturbative computation that the conformal symmetry is indeed preserved at one-loop provided the SI scheme is used, i.e. to verify that the counterterms can be put in a conformally symmetric form.

%The main claim 
%of the present paper 
%is that for a theory in flat space time the special conformal transformation defined by
%\be
%\Phi ' (x) = \f {1} {(1- 2 (a x) + a ^ 2 x ^ 2)^ {\D}} \Phi \l ( \f {x - a x^2} {1- 2 (a x) + a ^ 2 x ^ 2} \r ),
%\ee
%where $a$ is a four-vector (parameter of the transformation) and $\Delta$ is the scaling dimension of the field $\Phi$, survives at the quantum level
 %-- meaning that the conformal Ward identities are not anomalous -- 
 %if one uses a {\it regularization} consistent with the symmetry. For the dimensional regularization it amounts to say that the {\it regularized} Lagrangian is invariant under the conformal transformations.

%This is one of the reasons we consider only flat space-time leaving the study of systems coupled to gravity for future investigations.

We also consider the problem from a different perspective. In classical physics according to the N\"other theorem there is a conserved current corresponding to each continuous symmetry. In quantum field theory (QFT) the manifestation of a symmetry is the Ward identity involving N\"other currents. In the case of the conformal group one can build fifteen conserved currents
\be
J ^ a _\mu = T_{\mu \nu} \, \epsilon ^\nu _ a,
\ee
where $T _ {\m \n}$ is the energy-momentum tensor (EMT) and $\eps ^ \m$ is the parameter of the corresponding space-time transformations
\be
x'^\mu \rightarrow x^\mu + \epsilon^\mu.
\ee
In particular, the invariance under special conformal transformations implies that the EMT is traceless \cite{Polchinski:1987dy}.
%symmetry if we consider an infinitesimal transformation $x'^\mu \rightarrow x^\mu + \epsilon^\mu$, then the associated conserved current $J_\mu$ can be written in terms of the EMT as $J_\mu = T_{\mu \nu} \, \epsilon ^\nu$. If the EMT is symmetric and conserved we have
%\be
%\partial ^{\mu} J_{\mu} =0 \qquad \rightarrow \qquad \frac{1}{d} \, T^\mu_\mu \, (\partial \cdot \epsilon)=0 
%\ee
%and we can see that in a conformal invariant theory the EMT has to be traceless, $T^\mu_\mu = 0$. 
%Therefore, the EMT plays a very important role and it can be used as a double-check of our procedure.

For the two analyzed toy models we will show that using the SI scheme it is possible to build  a traceless renormalized EMT
\be
\eta ^ {\m \n} \la [T _ {\m \n}] \dots \ra = 0,
\ee
meaning that conformal symmetry is preserved.
%Another remark concerns the form of the Green's functions considered in  this paper. We discuss Ward identities with only one insertion of the current. Taking into account multiple insertions of the composite operators such as the  EMT leads to the already mentioned Weyl anomaly.
The conformal Ward identities considered in this paper have only one insertion of the current, which is enough to make a conclusion about the trace anomaly in flat space-time. Taking into account multiple insertions of composite operators such as the EMT leads to the already mentioned Weyl anomaly, which is not discussed here.

The paper is organized in the following way: in Section 2 the SI prescription is described in detail and the way in which the renormalization scale is replaced by dynamical fields is clarified; Section 3 shows how to build a conserved traceless EMT for a scale invariant Lagrangian at the \emph{classical level}, so that the theory is conformally invariant; Section 4 and 5 illustrate our main results in two toy-models in 4 and in 6 dimensions respectively; in Section 6 we briefly present a generalization of method for the case of spinor and vector fields and we touch on the UV completion problem of these theories; Section 7 contains our conclusions.

\section{SI prescription}

In this section we start by first describing the standard way of proceeding in dimensional regularization and then the different approach provided by the use of the SI prescription~\cite{Shaposhnikov:2008xi,Shaposhnikov:2008xb}. To better illustrate the technique outlined in the introduction, we consider a model with one real scalar field and a Higgs-like potential
\be
\mc{L} = \f {1} {2} \l ( \p _ \m \phi _ 0 \r ) ^ 2 - \lambda _ 0 \l ( \phi _ 0^ 2 -  \la \phi _ 0 \ra ^ 2 \r ) ^ 2.
\label{Higgs_Lagr}
\ee
In order to make the system scale invariant at the classical level the dilaton $X$ is introduced. In general the renormalizable scale invariant Lagrangian for two scalar fields has the following form
\be
V_g = \lambda _ 0 \phi _ 0 ^ 4 + \lambda _ 0 ' X _ 0 ^ 4 + \lambda ^ {''} _ {0} \phi _ 0 ^ 2 X _ 0 ^ 2 = 
\lambda _ 0 \l ( \phi _ 0^ 2 - \z _ 0 ^ 2 X _ 0 ^ 2 \r ) ^ 2 + \b ' X _ 0 ^2.
\ee 
For $\b' < 0$ there is no ground state, therefore, this case is not considered. For $\b ' > 0$ in turn the ground state is unique $\phi_0 = X_ 0 = 0$. Such a state does not break scale symmetry, therefore, the theory does not have asymptotic states and hence, it does not allow for any particle interpretation. As a result one is forced to consider only the case with $\b'=0$  to have a phenomenologically viable theory.

As a result we are left with the following scale/conformal extension of (\ref{Higgs_Lagr})
\be
\mc{L} = \f {1} {2} \l ( \p _ \m \phi _ 0 \r ) ^ 2 + \f {1} {2} \l ( \p _ \m X _ 0 \r ) ^ 2  
- \lambda _ 0 \l ( \phi _ 0^ 2 - \z _ 0 ^2 X _ 0 ^ 2 \r ) ^ 2,
\label{mod_Higgs}
\ee
We will assume that the symmetry breaking scale $\la X \ra = v$ will be taken of order of the Planck mass $M_ P = 10 ^ {19}~\text{GeV}$. In this case the low energy physics -- for energies $E^2 \ll v^2$ -- is given by (\ref{Higgs_Lagr}) provided the coupling $\z$ is chosen to reproduce the vev of the Higgs $\la \phi \ra = \z v$, hence $\z \lll 1$.

The standard dimensional regularization demands that for a system described by the Lagrangian (\ref{mod_Higgs}) one introduces the renormalized dimensionless coupling constants $\lambda, \z$ in the following way
\bea
\lambda _ 0 & = & \m ^ {4-n} \l ( \lambda + \sum _ k \f {C _ k (\lambda,\z)} {(n-4) ^ k} \r ), \nn \\
\z _ 0 & = & \l ( \z + \sum _ k \f {D _ k (\lambda,\z)} {(n-4) ^ k} \r ),
\label{st_dim_reg}
\eea
where the sum contains all the divergent terms order by order in perturbation theory.
The parameter $\m$ is an arbitrary scale needed to compensate for the dimensionality of $\lambda _ 0$ in other than $4$ dimensions. It plays a crucial role being the only source of explicit symmetry breaking.

Now let's turn to the SI prescription. In this case the arbitrary scale $\mu$ is promoted to be a field dependent quantity
\be
\m ^ {4-n} \to \l (X ^ 2\r ) ^ {\f{4-n}{n-2}} F _n \l ( \phi / X \r ),
\label{mod_dim_reg}
\ee
with an arbitrary function $F_ n$ such that $F_ 4 = 1$. It is straightforward to check that scale and conformal symmetries become manifest, once the expression (\ref{mod_dim_reg})
%\be
%\mu ^ {4-n} \to \l (  \x _ \phi \phi _ 0 ^ 2 + \x _ X X _ 0 ^ 2 \r ) ^ \f {4-n} {n-2},
%\label{mod_dim_reg_spec}
%\ee
is substituted into~(\ref{mod_Higgs})
%\be
%\lambda _ 0 \to \lambda _ 0 \l (  \x _ \phi \phi ^ 2 + \x _ X X ^ 2 \r ) ^ \f {4-n} {n-2},
%\ee
%to the Lagrangian (\ref{mod_Higgs})
\be
\mc{L} = \f {1} {2} \l ( \p _ \m \phi _ 0 \r ) ^ 2 + \f {1} {2} \l ( \p _ \m X _ 0 \r ) ^ 2  
- \lambda _ 0 \l ( \phi _ 0^ 2 - \z _ 0 ^ 2 X _ 0 ^ 2 \r ) ^ 2 \l ( X _ 0 ^ 2 \r ) ^ {\f {4-n} {n-2}} F _ n.
\label{mod_Lagrangian}
\ee
The way to deal with the fractional power of the dilaton in the Lagrangian is to expand it around the vev
\be
X = v + \c.
\label{dilaton_vev}
\ee
As a result there is an infinite number of terms making the Lagrangian non-renormalizable~\cite{Shaposhnikov:2009nk}.

Computing the one-loop effective action, the authors of \cite{Shaposhnikov:2008xi} showed that it is scale invariant. At this point it may seem that this theory loses all contact with the real world. Usually the couplings in scale invariant theories do not run, since there is no quantity setting the scale. On the contrary, as all experimental evidences point out, couplings do run and this behavior is encoded in the renormalization group equations.

To clarify the situation we note that what is important is not the dependence of couplings on the renormalization scale, but rather the scaling of correlators (observables) with the energy of the process. For example in the limit $\z \lll 1$ and $\z v = \la \phi \ra \ll \sqrt{s} \ll v$ for a $2 \rightarrow 2$ Higgs scattering, the $4$-point function $\la \phi _ 1 \phi _ 2 \phi _ 3 \phi _4 \ra$ becomes
\be
\G _ 4 = \lambda  +  b _ 1 \lambda ^ 2 \log \f {s} { v ^ 2} + O (\z ^ 2),
\label{lambda_run}
\ee
%\f {9 \lambda ^ 2} {64 \pi ^ 2}
where $b_1$ is the first coefficient of the expansion of the $\b$-function
\be
\b(\lambda) = b_1 \lambda ^ 2 + b_2 \lambda ^ 3 + \dots.
\ee
One sees that the expression (\ref{lambda_run}) coincides with the one prescribed by the ordinary renormalization group.

The result (\ref{lambda_run}) heavily relies on the spontaneous breaking of the scale invariance. Therefore, in order for the whole approach to be consistent, the flat direction providing the spontaneous symmetry breaking has to be preserved. If it does not survive quantum corrections, the expansion around the chosen ground state becomes inconsistent because it is not the vacuum anymore. The authors of \cite{Shaposhnikov:2008xi} showed that in fact the flat direction of the effective potential can be consistently preserved by adding scale invariant finite counterterms.

%The observation that one can make is that such a modification of the Lagrangian automatically renders it conformally symmetric.
%In \cite{Fubini:1976jm} classically conformally invariant Lagrangian is considered in arbitrary number of dimensions (potential does not have symmetry breaking vacua, only zero). Spontaneous breaking of the conformal symmetry is considered. It is shown that the ground state is not translationally invariant, however, it can be invariant under a 10-dimensional subgroup of the conformal $O(n,2)$ group. Translational invariance is reached only upon averaging over all degenerate vacua (thermodynamics).

\section{Classical energy-momentum tensor}

Now we turn to the subject of our study. We consider a system of two interacting scalar fields described in the previous section. Building the energy momentum tensor and renormalizing it at one-loop level we show that it stays traceless, provided one works within the SI prescription. Although for the explicit computation purpose we are going to choose a specific form of the Lagrangian, it is instructive to get the expression for the classical EMT for an arbitrary scale invariant potential. The system is described by the following classical Lagrangian
\be
\mc{L} = \f {1} {2} \l ( \p _ \m \phi _ 0 \r ) ^ 2 + \f {1} {2} \l ( \p _ \m X _ 0 \r ) ^ 2  - V (\phi_0,X_0),
\label{arb_Lagr}
\ee
where the potential $V$  is chosen in such a way that the Lagrangian is scale invariant in an arbitrary number of 
dimensions $n$. It means that the potential necessarily depends on $n$ and it is of degree $\f {2n} {n-2}$
\be
\l ( \phi _ 0 \f {\p} {\p \phi _ 0} + X _ 0 \f {\p} {\p X _ 0} \r ) V = \f {2n} {n-2} V.
\ee
The system is automatically conformally invariant at the classical level. Indeed, the improved EMT is given by
\be
T _ {\m \n} = \p _ \m \phi _ 0 \p _ \n \phi _ 0 + \p _ \m X _ 0 \p _ \n X _ 0 - g _ {\m \n} \mc{L} -
\xi _ C \l ( \p _ \m \p _ \n - g _ {\m \n} \p ^ 2 \r ) \l [ \phi _ 0 ^ 2 + X _ 0 ^ 2 \r ],
\label{EM_gen}
\ee
with
\be
\xi _ C = \f {1} {4} \f {n - 2} {n - 1}.
\ee
It proves helpful to introduce the following tensor 
\be
t _ {\m \n} = n \p _ \m \p _ \n - \eta _ {\m \n} \p ^ 2,
\ee
traceless in an arbitrary number of dimensions, and the notations for the operators
\bea
E _ \phi ^ 0 & = & \p ^ 2 \phi _ 0 + \f {\p V} {\p \phi  _ 0}, \nn \\
E _ X ^ 0 & = & \p ^ 2 X _ 0 + \f {\p V} {\p X  _ 0}. \nn
\eea
As a result one gets the EMT in the following form
\be
T _ {\m \n} = \f {1} {4} \f {1} {n - 1} t _ {\m \n} \l ( \phi _ 0 ^ 2 + X _ 0 ^ 2 \r ) - 
\f {1} {n} \l ( X _ 0 t _ {\m \n} X _ 0 + \phi _ 0 t _ {\m \n} \phi _ 0 \r ) 
+ \eta _ {\m \n} \f {n-2} {2n} \l ( \phi _0 E _ \phi ^ 0 + X _ 0 E _ X ^ 0 \r ).
\label{EM_gen_t}
\ee
It is straightforward to show that at the classical level such an EMT is traceless up to contact terms. Therefore, the system is conformally invariant in an arbitrary number of dimensions. Our goal is to check whether this feature survives one-loop radiative corrections.

\section{$4$-dimensional case}

In this section we turn to the explicit computation carried out in the first one of the two toy models to demonstrate how one deals with the prescription described above. For the sake of simplicity instead of considering the Lagrangian (\ref{mod_Higgs}) we consider its limit corresponding to $\z=0$ and we choose the function $F _ n = 1$ to get the Lagrangian in the form
\be
\mc{L} = \f {1} {2} \l ( \p _ \m \phi _ 0 \r ) ^ 2 + \f {1} {2} \l ( \p _ \m X _ 0 \r ) ^ 2  
- \f{\lambda _ 0} {4!} \phi _ 0 ^ 4 \l ( X _ 0 ^ 2 \r ) ^ {\a_4}.
\label{n4_Lagrangian}
\ee

%To this end we use the toy-model Lagrangian  which is simply the conformal extension of a $4$-dimensional $\phi ^4$ theory to an arbitrary number of dimensions
%\be
%\mc{L} = \f {1} {2} \l ( \p _ \m \phi _ 0 \r ) ^ 2 + \f {1} {2} \l ( \p _ \m X _ 0 \r ) ^ 2  
%- \f{\lambda _ 0} {4!} \phi _ 0 ^ 4 \l ( X _ 0 ^ 2 \r ) ^ {\a_4},
%\label{n4_Lagrangian}
%\ee
%where $\a _ 4 =( {4-n} ) / ( {n-2} )$ and the index $0$ means that the corresponding quantity is a bare one.

%
%
%
%
%
%The Lagrangian at hand is not renormalizable \cite{Shaposhnikov:2009nk} (we will define in a moment how one should deal with the theory when $\f {2n} {n-2}$ is fractional). Therefore, the outline of our approach is as follows. We renormalize the theory at a fixed order in perturbation theory (here we will not consider any corrections beyond the one loop approximation) by choosing the appropriate counterterms needed to cancel the  divergences in the Green's functions  involving fundamental fields. Then we build the EMT, renormalize it and show that it stays traceless.

%Classically the Lagrangian above is both scale and conformally invariant. In order to quantize the system we choose the vacuum and expand the fields around it
%\bea
%\phi_ 0 & \to & \phi _ 0, \nn \\
%X & \to & v _ 0 + \c _ 0.
%\eea
%
%%
%%
%%
%%
%%
%%
%%
%%
%%
%%
%
%
%As a result we have
%\be
%\mc{L} = \f {1} {2} \l ( \p _ \m \phi _ 0 \r ) ^ 2 + \f {1} {2} \l ( \p _ \m \c _ 0 \r ) ^ 2  
%- \f{\lambda _ 0 v _ 0 ^ {2 \a _ 4}} {4!} \phi _ 0 ^ 4 \l ( 1 + \f {\c _ 0} {v _ 0} \r ) ^ {2 \a _ 4}.
%\label{4n_Lagr}
%\ee
Once expanded in powers of $\a _ 4 =( {4-n} ) / ( {n-2} )$, keeping in mind that the dilaton acquires the vev (\ref{dilaton_vev}), the Lagrangian becomes
\be
\mc{L} = \f {1} {2} \l ( \p _ \m \phi _ 0 \r ) ^ 2 + \f {1} {2} \l ( \p _ \m \c _ 0 \r ) ^ 2  
- \f{\lambda _ 0 v _ 0 ^ {2 \a _ 4}} {4!} \phi _ 0 ^ 4 \l \{ 1 + 2 \a _ 4 \log \l ( 1 + \f {\c _ 0} {v _ 0} \r ) + O (\a _ 4 ^ 2) \r \}.
\ee
As $\a _ 4$ is proportional to $n-4$, we see that at the one-loop level there are no divergent diagrams containing vertices with a $\c$ field and therefore the structure of divergences of this theory is the same as for the $\phi ^ 4$ theory.

\begin{figure}[H]
\centering
\includegraphics[width=6cm]{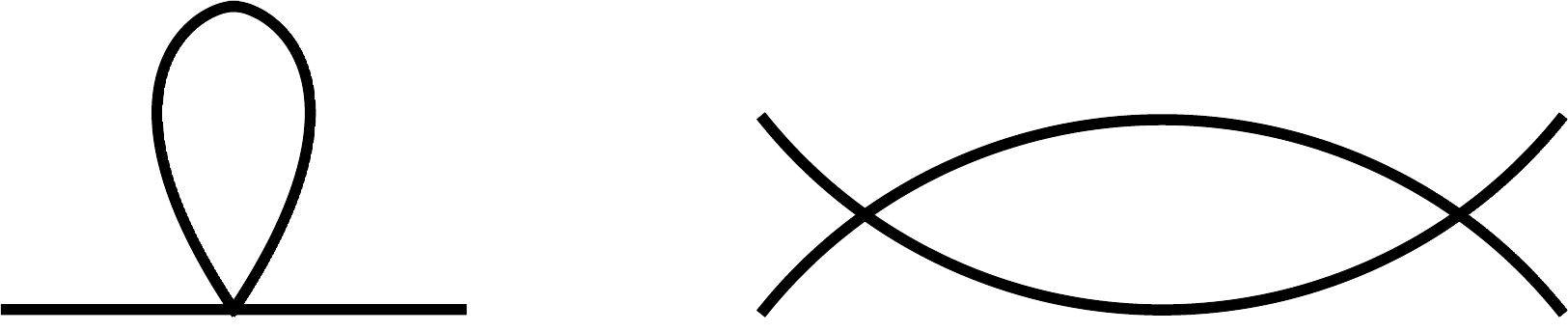}
\caption{One-loop divergent diagrams in a $\phi ^ 4$ theory.\label{n4_1loop}}
\end{figure}

The only  one-loop diagrams to be renormalized are the ones shown in Fig.~\ref{n4_1loop} (in fact there are three diagrams with four external legs corresponding to the $s,t$ and $u $ channels). In dimensional regularization the first diagram may be put to zero for  a massless theory (there is no field renormalization, hence, $\phi _ 0 = \phi$,~$\c _ 0 =\c$), while to make finite the second diagram we can take the result from $\phi ^ 4$ theory. The counterterm
\be
\f {3} {16 \pi^2} \f {\lambda ^ 2 \, v ^ {2 \a _ 4}} {4-n} \f {\phi ^ 4} {4!},
\label{cterm}
\ee
takes care of the divergence, but such a term is not conformally invariant {\it per se}. However, one notes that the one-loop counterterm  in (\ref{cterm})  is nothing but the first term of the expansion 

\be
\f {1} {4-n} \f {3 \lambda  ^ 2} {16 \pi^2 }\f {\phi ^ 4} {4!} \l ( X ^ 2 \r ) ^ {\a _ 4} 
\simeq
\f {3} {16 \pi^2} \f {\lambda  ^ 2 \, v  ^ {2 \a _ 4}} {4-n} \f {\phi ^ 4} {4!} \l \{ 1 + 2 \a _ 4 \log \l ( 1 + \f {\c } {v } \r ) + O (\a _ 4^2) \r \},
\label{conf_cterm}
\ee 
where the structure in the {\it l.h.s.} of (\ref{conf_cterm}) is conformally invariant. 
%\footnote{It corresponds to 
%$\f {1} {4-n} \f {3 \lambda _ 0 ^ 2} {16 \pi^2 }\f {\phi ^ 4} {4!} \l ( X ^ 2 \r ) ^ {\a _ 4}$ which is conformal.})
%\be
%\f {3} {16 \pi^2} \f {\lambda _ 0 ^ 2 \, v _0 ^ {2 \a _ 4}} {4-n} \f {\phi ^ 4} {4!} \l \{ 1 + 2 \a _ 4 \log \l ( 1 + \f {\c _ 0} {v _ 0} \r ) + O (\a _ 4^2) \r \}.
%\ee
Adding this counterterm amounts to the renormalization of the coupling constant
\be
\lambda _ 0 = \lambda + \f {3} {16 \pi^2} \f {\lambda ^ 2 } {4-n}.
\label{lambda_1loop}
\ee
As a result we have that the Lagrangian renormalized at one loop has the same form as the one in (\ref{n4_Lagrangian}) with renormalized fields and coupling constant given by (\ref{lambda_1loop}). Therefore, using (\ref{EM_gen_t}) and the results of 
%we get the expression for the  EMT in the form
%\be
%T _ {\m \n} = \f {1} {4} \f {1} {n - 1} t _ {\m \n} \l ( \phi ^ 2 + \chi ^ 2 \r ) - 
%\f {1} {n} \l ( \c t _ {\m \n} \c + \phi t _ {\m \n} \phi \r ) -
%\f {2v \xi _ C} {n} t _ {\m \n} \c + \f {1} {n} \eta _ {\m \n} \l ( \f {n} {2} - 1 \r ) \l \{ \phi E _ \phi + (v + \c) E _ \c \r \}.
%\ee
%Now one can use the results 
\cite{Brown:1979pq} one can show that the EMT is finite\footnote{For the renormalization of composite operators see \cite{Collins_book}.} and traceless
\be
\l [ T _ {\m \nu} \r ] \eta ^ {\m \n} = 0.
\ee

Although for the illustration purposes we used a specific limit of the Lagrangian (\ref{mod_Higgs}), we do not expect the conclusion reached now about the conformal invariance to change if we consider a generic value for $\z$ and an arbitrary function $F _ n$.

%
%Secondly, wasn't it just a coincidence since for the toy model we have chosen renormalizable (within the standard approach) $\phi ^ 4$ theory and extended it?

\section{$6$-dimensional space}

The explicit example of the previous section shows that the conformal symmetry of the system can be maintained at the one-loop level using the SI prescription. One might raise  some objections. Firstly at the one-loop level there is no correction other than the renormalization of 
$\lambda$, for the toy model that we have chosen is renormalizable within the standard approach. As a result the form of the Lagrangian is not affected by the introduction of the counterterms and hence can be made conformal. It might well happen that once multi loop corrections are accounted for, new symmetry breaking terms are generated in the Lagrangian.  Secondly, since there are no virtual dilatons propagating in the loops, the result should be obvious since in this case the modification is effectively equivalent to the one discussed in~\cite{Komargodski:2011xv,Komargodski:2011vj,Luty:2012ww,Codello:2012sn}.

We believe that it is possible to show that the system stays conformal even at a higher number of loops, where nonrenormalizability and propagating dilatons come into play. However, to simplify the technical side of the problem we present below an example where the mentioned effects appear already at one loop.

\begin{figure}[H]
\centering
\includegraphics[height=3cm]{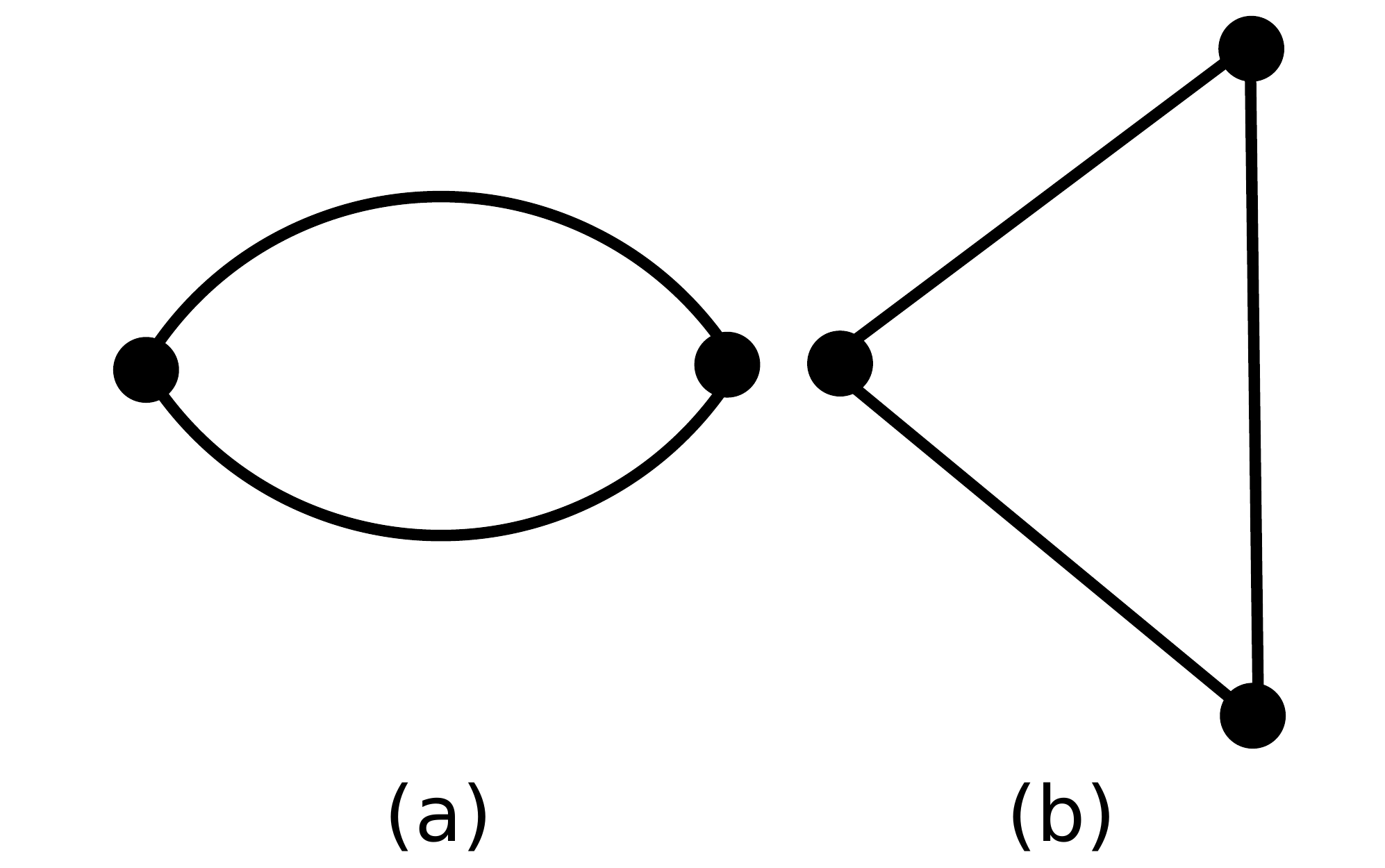}
\caption{One-loop divergent graphs with topology of the \emph{bubble} (a) and \emph{triangle} (b) in a $6$ dimensional scalar theory. \label{top_div}}
\end{figure}

Let us consider a system  of two scalar fields, but this time in 6 dimensions. Its Lagrangian  is
\be
\mc{L} = \f {1} {2} \l ( \p _ \m \phi _ 0 \r ) ^ 2 + \f {1} {2} \l ( \p _ \m X _ 0 \r ) ^ 2  
- \f{\lambda _ 0} {4!} \f{\phi _ 0 ^ 4} {X _ 0} \l ( X _ 0 ^ 2 \r ) ^ {\a_6},
\label{n6_Lagrangian}
\ee
where $2 \a _ 6 = {( 6-n )} /{(n-2)}$. Since $\a _ 6 \rightarrow 0$ (vanishes) as $n \rightarrow 6$ the terms $O (\a _ 6)$ do not contribute at one loop and, therefore, can be dropped. As before the term $X _ 0 ^ {-1} = (v _ 0 + \c _ 0 ) ^ {-1}$ is to be understood as a series in powers of the ratio ${\c _ 0} / {v _ 0}$. Expanding the potential in (\ref{n6_Lagrangian}) we get
\be
V(\phi_ 0, \c _ 0) =  \f{\lambda _ 0 v _ 0 ^ {2 \a _ 6}} {4! v _ 0} \phi _ 0 ^ 4 \l \{ 1 - \f {\c _ 0} {v _ 0} + \l ( \f {\c _ 0} {v _ 0} \r ) ^ 2 
+ O \l ( \f {\c _ 0 ^ 2} {v _ 0 ^ 3} \r ) \r \},
\label{potential_exp}
\ee
which leads to the Feynman rules for all the vertices with an increasing number of $\c _ 0$ legs. The topology of the divergent one-loop graphs is depicted in Fig.~\ref{top_div}.

\begin{figure}[H]
\centering
\includegraphics[width=6cm]{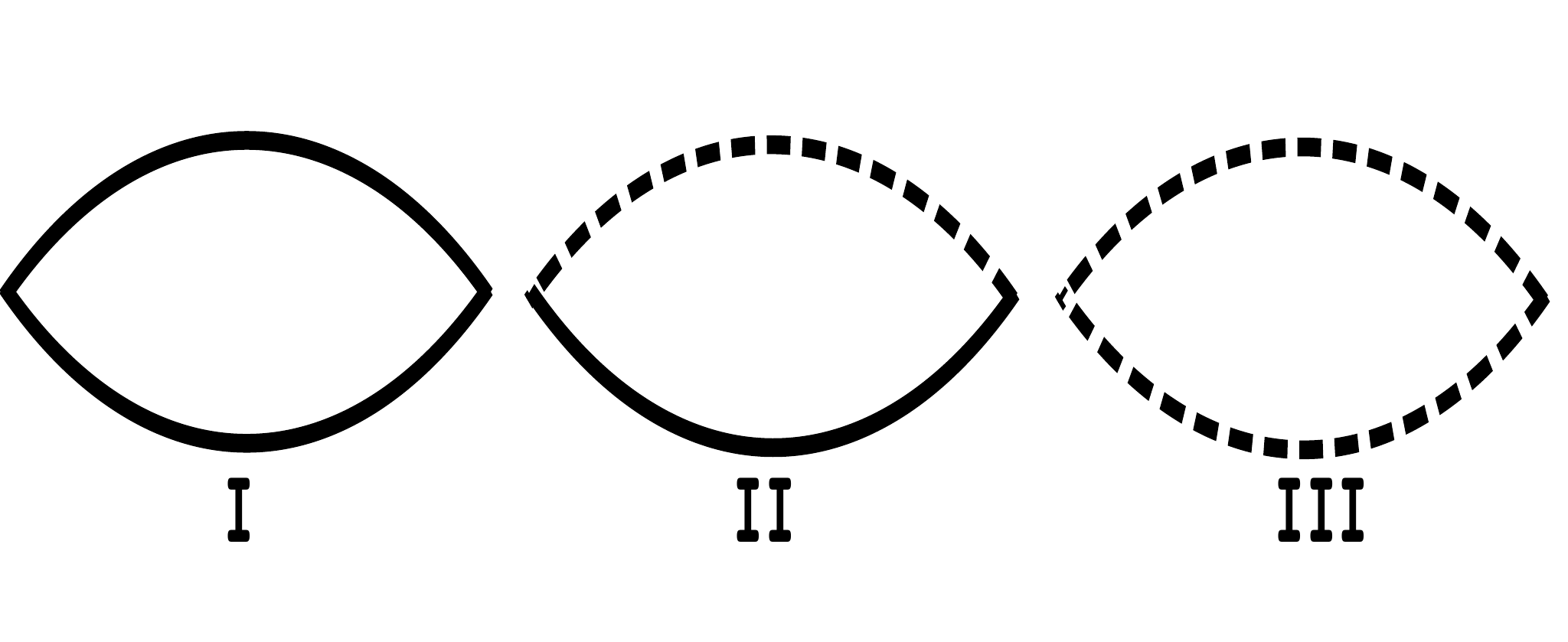}
\caption{One-loop diagrams of the \emph{bubble} topology in which  $\phi$ and $\c$ fields are drawn as solid and dashed lines correspondingly.
\label{n6_1loop_top} }
\end{figure}

The triangular graphs of the type in Fig.~\ref{top_div}b are logarithmically divergent, therefore, they correspond to the introduction of potential-like counterterms. It seems rather obvious that such counterterms can  always be made conformal at one loop if multiplied  by an appropriate power of $X = v + \c$, which amounts to a finite renormalization. On the other hand the graphs of the type in Fig.~\ref{top_div}a are quadratically divergent, therefore, they correspond to new counterterms with two derivatives. Three distinct kinds of loops with different fields on the internal lines have the same \emph{bubble} topology in Fig.~\ref{top_div}a. They are all shown in Fig.~\ref{n6_1loop_top}.

\begin{figure}[H]
\centering
\includegraphics[height=5cm]{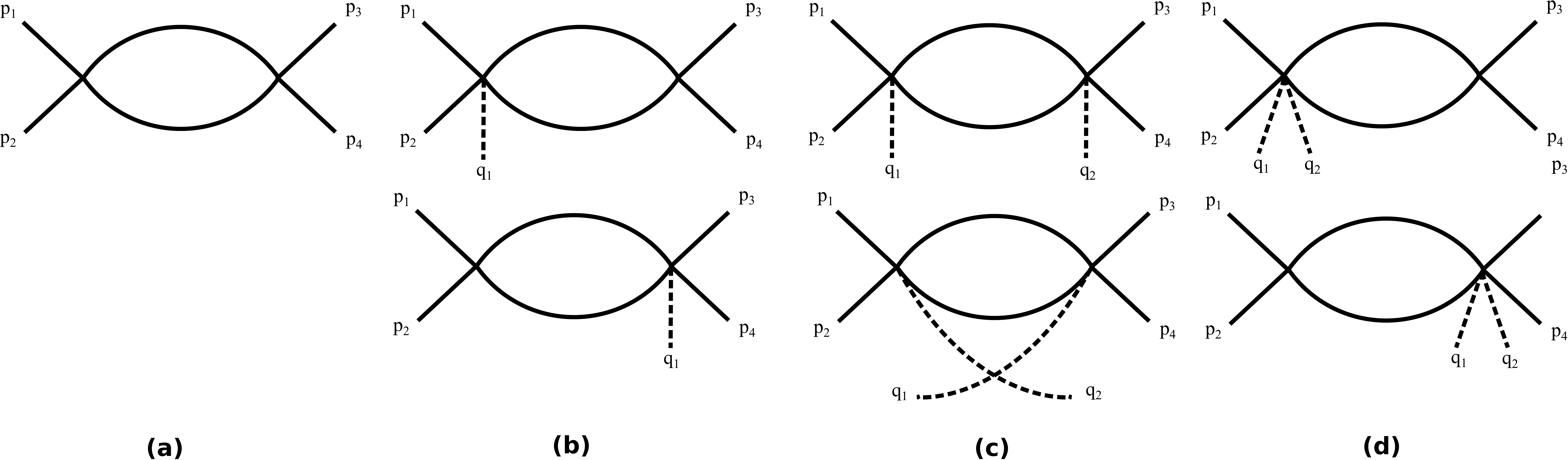}
\caption{First terms of the expansion in $v ^{-1}$. To each $\c$ insertion (dashed line) corresponds an additional power of $v ^{-1}$.\label{n6_1loop_v}}
\end{figure}

Let us first consider the diagrams from the group I in which the $\phi$ field runs in the loop and the field $\c$ may only appear on the external legs. The three leading terms of the expansion in $v ^ {-1}$ are depicted in Fig.~\ref{n6_1loop_v}. 

Introducing the following notation for the loop integral with two scalar propagators
\be
F _ n ( p ) = \int \f {d ^ n k} {(2 \pi) ^ n} \f {1} {k^2 (k+p) ^ 2} = \f {i} {(4 \pi) ^ {n / 2}} (- p ^ 2) ^ {\f {n} {2} - 2}
\f {\G (2 -  {n} / {2}) \G ^ 2 ( {n} / {2} - 1)} {\G ( {n} - {2})}
\label{function_F}
\ee
and taking into account all three channels (configurations for $\phi$ legs) we obtain the following result for the diagram in the first column of Fig.~\ref{n6_1loop_v}
\be
\text{(a)} = i  \f {\lambda ^ 2 v ^ {4\a_6}} {4 v ^ 2}
\l \{ F _ n (p _ {12}) + F _ n (p _ {13}) + F _ n (p _ {14}) + F _ n (p _ {23}) + F _ n (p _ {24}) + F _ n (p _ {34}) \r \},
\label{6n_ct_a}
\ee
where $p _ {ij} = p_ i + p _ j$, and we use the convention that all the momenta are incoming. Similarly one finds for the remaining diagrams
\bea
\text{(b)} & = & - i \f { \lambda ^ 2 v ^ {4\a_6}} {2 v^3}
\l \{  F _ n (p _ {12}) + F _ n (p _ {13}) + F _ n (p _ {14}) + F _ n (p _ {23}) + F _ n (p _ {24}) + F _ n (p _ {34}) \r \}, \nn \\
\text{(c)} & = & i \f { \lambda ^ 2 v ^ {4\a_6}} {2 v ^ 4}
\l \{  F _ n (p _ {121}) + F _ n (p _ {131}) + F _ n (p _ {141}) + F _ n (p _ {122}) + F _ n (p _ {132}) + F _ n (p _ {142}) \r \}, \nn \\
\text{(d)}& = & i \f {\lambda ^ 2 v ^ {4\a_6}} {v ^ 4}
\l \{  F _ n (p _ {12}) + F _ n (p _ {13}) + F _ n (p _ {14}) + F _ n (p _ {23}) + F _ n (p _ {24}) + F _ n (p _ {34}) \r \},
\label{6n_ct_bcd}
\eea
with $p _ {ija} = p _ i + p _ j + q _ a$. The structure of the divergences precisely coincides with the expansion of the following counterterm
\be
\mc{L} _ {1} = \f {1} {(2!)^2} \f{ C _ 1} {6 - n} \, \f {\phi ^ 2} {X} \p ^ 2 \f {\phi ^ 2} {X}.
\label{6n_ct}
\ee
Namely,
\bea
(\text{a}) & \to & \f{ 1 } {6 - n} \,\f {C _ 1} {v ^ 2} \phi ^ 2 \p ^ 2 \phi ^ 2, \nn \\
(\text{b}) & \to & - 2 \f{ 1 } {6 - n} \, \f {C _ 1} {v ^ 2} \f {\c} {v} \phi ^ 2 \p ^ 2 \phi ^ 2 , \nn \\
( \text{c} ) & \to & \f{ 1 } {6 - n} \, \f {C _ 1} {v ^ 2} \, \f {\c} {v} \phi ^ 2 \p ^ 2 \f {\c} {v} \phi ^ 2, \nn \\
( \text{d} ) & \to & 2 \f{ 1 } {6 - n} \, \f {C _ 1} {v ^ 2} \f {\c ^ 2} {v ^ 2} \phi ^ 2 \p ^ 2 \phi ^ 2,
\eea
with
\be
C _ 1 = \f {\lambda ^ 2} {12 (4\pi)^3}.
\label{C1_coeff}
\ee
Therefore, we have demonstrated that up to $v^{-4}$ order the counterterm (\ref{6n_ct}) with the coefficient $C_1$ given by (\ref{C1_coeff}) cancels the divergence. In the Appendix we show that this is in fact the case for an arbitrary order in $v$. Similarly, one proves that for the diagrams of types II and III in Fig.~\ref{n6_1loop_v} the following counterterms are needed
\bea
\mc{L} _ {2} = \f {1} {(3!)^2} \f {C _ 2} {6-n} \,\f {\phi ^ 3} {X^2} \p ^ 2 \f {\phi ^ 3} {X^2}, \nn \\
\mc{L} _ {3} = \f {1} {(4!)^2} \f {C _ 3} {6-n} \, \f {\phi ^ 4} {X^3} \p ^ 2 \f {\phi ^ 4} {X^3},
\label{6n_ct_II}
\eea
with
\bea
C _ 2 & = & \f {\lambda ^ 2} {6 (4\pi)^3}, \nn \\
C _ 3 & = & \f {\lambda ^ 2} {3 (4\pi)^3}.
\label{C23}
\eea

It is straightforward to show that the counterterms $\mc{L}_{1,2,3}$ are conformally invariant. Their presence in the Lagrangian renders it sigma model-like
%\be
%\mc{L}_ {\text{1-loop}} = g _ {i j} \p _ \m \phi ^ i \p _ \m \phi ^ j - V,
%\ee
%where $\phi _ 1 = \phi$ and $\phi _ 2 = X$ and the 
with  a flat metric. Therefore, one can make the coordinate transformation 
\be
(\phi , X) \mapsto (\phi', X'),
\ee
to recast the kinetic term in the canonical form. As a result, we reduced the problem to the one discussed in the previous section. Therefore, the renormalized EMT is traceless.

\section{Discussion}

\subsection{Including vectors and fermions}

Although in the paper we focused on theories with only scalar fields and we demonstrated only for the scalar fields that the SI prescription works, the technique can be easily generalized to include vector and spinor fields as well. The main idea is to make the regularized Lagrangian conformally invariant. Using dimensional regularization it consists of extending the Lagrangian to an arbitrary number of dimensions in a conformally invariant way. One should only take care of the terms in the Lagrangian whose couplings become dimensionful when in dimensions $n$ other than $4$. It is rather clear that for any field theory renormalizable in a standard way, the one-loop result is going to be conformally invariant if the SI prescription is used. The reason is the following. All the couplings in $4$ dimensions in these theories can be written as non negative integer powers of the dilaton field. The modification one introduces in $n$ dimensions is that the powers of the dilaton field acquire corrections proportional to $n-4$. These corrections can be neglected at the one-loop order for they do not change the structure of the divergences. The resulting counterterms can be made conformal by adding finite renormalization.

To demonstrate it we take into account the Lagrangian for quantum electro-dynamics (QED). This choice allows us to show how the framework can be also applied to theories involving fields other than scalars.

The standard Lagrangian for QED in $4$ dimensions
\be
\mc L _ {QED} = - \f {1} {4 e ^ 2} F _ {\m \n} ^ 2 + i \bar \psi \g ^ \m \l ( \p _ \m - i A _ \m \r ) \psi - m \bar \psi \psi,
\label{QED_st}
\ee
can be modified in the following way
\bea
\mc L ^ {reg} _ {QED} & = & - \f {1} {4 e ^ 2} F _ {\m \n} ^ 2  ( X ^ 2 ) ^  {\f {n-4} {n-2}} + i \bar \psi \g ^ \m \l ( \p _ \m - i A _ \m \r ) \psi + \f {1} {2} \p _ \m X \p ^ \m X - g X \bar \psi \psi ( X ) ^ {\f {4-n} {n-2}} - \lambda X ^ 4  \label{QED_mod} \\
& = & - \f {1} {4 e ^ 2} F _ {\m \n} ^ 2  ( v ^ 2 ) ^  {\f {n-4} {n-2}} + i \bar \psi \g ^ \m \l ( \p _ \m - i A _ \m \r ) \psi + \f {1} {2} \p _ \m \c \p ^ \m \c - g v ^ {\f {4-n} {n-2}}   v \bar \psi \psi - g v ^ {\f {4-n} {n-2}} \c \bar \psi \psi - \lambda X ^ 4, \nn
\eea
where in the second line we expanded around $\la X \ra = v$, keeping only the terms relevant at the one-loop level. In (\ref{QED_mod}) one recognizes the standard QED Lagrangian (\ref{QED_st}) with a fermion mass $m=gv$ and an additional Yukawa interaction $\c \bar \psi \psi $. The renormalization of the Lagrangian (\ref{QED_mod}) leads to the inclusion of the counterterms
\bea
\mc L _ {QED} ^ {c.t.} & = &  - \f {\d _ e} {4 e ^ 2} F _ {\m \n} ^ 2  ( v ^ 2 ) ^  {\f {n-4} {n-2}} + i \d _ 2 \bar \psi \g ^ \m \l ( \p _ \m - i A _ \m \r ) \psi + \f {\d _ \c} {2} \p _ \m \c \p ^ \m \c - \d _ g v ^ {\f {4-n} {n-2}} ( v + \c ) \bar \psi \psi \nn \\
& - & \l ( \f {\d _ m} {2}  v ^ 2 X ^ 2 
+ \f {\d _ 3} {3!} v X ^ 3 + \f {\d _ 4} {4!} X ^ 4 \r ) v ^ {2 \a _ 4} ,
\eea
where $\a _ 4 = \f {4-n} {n-2}$, with $X = v + \c$, while $\{\d\}_a$ are dimensionless numbers (containing the pole structure $\f {1} {n-4}$). One sees that the counterterm Lagrangian can be written in a conformally invariant form by adding finite renormalization
\bea
\mc L _ {QED} ^ {c.t.} & = &  - \f {\d _ e} {4 e ^ 2} F _ {\m \n} ^ 2  ( X ^ 2 ) ^  {\f {n-4} {n-2}} 
+ i \d _ 2 \bar \psi \g ^ \m \l ( \p _ \m - i A _ \m \r ) \psi + \f {\d _ \c} {2} \p _ \m X \p _ \m X 
- \d _ g X ^ {\f {2} {n-2}} \bar \psi \psi \nn - \d _ {4X} \f {\d _ m} {4!}  X ^ 4.
\eea

As already mentioned the tracelessness of the renormalized EMT confirms that the theory stays conformally invariant at the quantum level. It is worth noticing in what the results obtained with the Lagrangians in  (\ref{QED_st})  and  (\ref{QED_mod}) differ. Using (\ref{QED_st}) even in the limit $m \to 0$ we find the non vanishing trace of the EMT
\be
\la   T _\m ^\m\ra  = - \f {e^2}{24 \pi^2} F_{\m \n}F ^{\m \n},
\ee
proportional to the $\b$-function of QED. This means the theory shows the presence of the trace anomaly \cite{Adler:1976zt,Giannotti:2008cv}. If we use modified QED Lagrangian (\ref{QED_mod}) instead, we find  $\la   T _\m ^\m\ra = 0$, therefore, the theory is conformally invariant. These two results are not in contradiction, as the matter content of the two theories is different --  a new degree of freedom, the dilaton, appears in  (\ref{QED_mod}).

\subsection{UV completion}

%Assuming that there indeed exists a theory with spontaneously broken conformal invariance having a fixed low energy limit, it is plausible to expect that with the formalism described in the previous section the low energy behavior of the theory can be reconstructed. This is precisely what we did up to now.

Up to now we have considered the perturbative expansion in the theory with spontaneously broken conformal symmetry, which is valid only in the low energy limit, for energies smaller then the symmetry breaking scale
\be
E ^ 2 < v ^ 2.
\ee
Increasing the energy we enter into a strongly coupled regime and the perturbative approach breaks down.
For energies of order of $v$, more and more terms become important and a resummation of all corrections is necessary  for determining the UV limit. 

In general the effective Lagrangian has an infinite number of free parameters corresponding to the finite parts of the counterterms. These parameters do not change the low energy dynamics, when the typical scale of momenta is much smaller then the vev of the dilaton $p ^ 2 \ll \la X \ra ^ 2$. However, they are crucial for the high energy behavior of the theory when $p^ 2 \sim \la X \ra ^ 2$. It looks conceivable that there may exist a set of these parameters, leading to a good behavior at energies $E \gg v$ corresponding to some CFT with unbroken symmetry.

If this indeed happens, there should be some relation between the low energy degrees of freedom, described by IR fields with canonical dimensions, and the high energy variables, in terms of which a CFT is formulated. The general way to describe a CFT is to provide the following set of data: the operators $\Phi_i$ with their dimensions $\D_i$ and the structure constants $C_{ijk}$~\cite{DiFrancesco:1997nk,Blumenhagen:2009zz}. The duality or correspondence can be formulated in this case as follows. All the operators of the UV theory $\Phi _ i$ can be written in terms of the fields present in the classical Lagrangian. Since all the fields by construction have no anomalous dimension, one can use simple dimensional analysis arguments to restrict the form of the correspondence. The field $\Phi$ with dimension $\D$ may include all possible combinations of fields from the classical Lagrangian with the corresponding dimension. For instance\footnote{The locality of the relation is not mandatory, non-local expressions are also possible. The only stipulation is the coherent transformation properties of both sides under the action of the whole conformal group.},
\be
\Phi \supset \l ( C _ X X + C _ \phi \phi + C _{\psi \psi} \f {\bar \psi \psi} {X^2} + \dots \r ) ^ {\D}, ~~C_ {\phi X} ( \phi X ) ^ {\f {\D} {2}}, ~~\text{etc.}
\label{conf_corr}
\ee
 The equality 
(\ref{conf_corr}) should be understood in the sense that the correlator
\be
\la \Phi _ 1 \dots \Phi _ N \ra,
\ee
known in the UV complete CFT, can be equally computed from the renormalized Lagrangian if one makes the identification (\ref{conf_corr}).

Clearly, the search for theories where the scenario described above is realized, is extremely challenging and it is not attempted here.

\section{Conclusion}

The key feature of the spontaneous symmetry breaking is the existence of the Goldstone mode. In the case of conformal symmetry breaking it is the dilaton. Therefore, the signature of a theory with spontaneously broken conformal symmetry is the presence of a massless scalar particle.

Although  the classical Lagrangian of a renormalizable field theory may possess conformal symmetry, taking into account radiative corrections usually reveals the non invariance of the system under scale (conformal) transformations. Due to the resulting running of the couplings, the Ward identities corresponding to the symmetries become anomalous. Therefore, in order to have a consistent picture of a spontaneously broken conformal symmetry (preserving the dilaton massless), a whole new approach should be developed. 

In the paper we investigated in more detail the SI scheme claiming that regularizing a theory in a scale invariant way would lead to the  presence of this symmetry even at the quantum level after renormalization.  For the dimensional regularization it amounts to say that the Lagrangian has to be modified to become scale invariant in an arbitrary number of dimensions. Such a modification leads to a non-renormalizable theory. This is not an obstacle as the final theory should include gravity.

The main observation is that the SI {\it regularization} can be made conformal. Our main conjecture is that in this case it is possible to {\it renormalize} the theory by adding {\it conformally invariant} counterterms. This means that the Lagrangian stays conformally invariant and one can build  a  traceless EMT. As a result the scale/conformal Ward identities are not anomalous. 

We showed that this is indeed the case at the one-loop level for the two toy models we considered. The first model is  a $4$-dimensional SI modified $\phi^4$ theory. We showed that at  the one-loop level the quantum corrections lead to the usual one-loop renormalization of the coupling constant, i.e. the counterterm is potential-like. Therefore, since the theory is scale invariant, it is automatically conformally invariant as well and hence, the EMT stays traceless. The second model given by the Lagrangian in (\ref{n6_Lagrangian}) is an example in $6$ dimensions. We showed that in this case there are sigma model-like counterterms. The metric in the space of the fields happens to be flat, therefore, the system is conformally invariant also in this case.

Although we expect this to be true for an arbitrary number of loops we did not present an explicit multi-loop computation  confirming that the system stays conformally invariant, nor can we give a general proof.

% Moreover, the presence of the term $X^{2\a}$ seems crucial in both cases considered in the paper (precisely this renders the classical potential conformally invariant in an arbitrary number of dimensions). However, the effect of such a term at one loop appears only in the definition of the unrenormalized  EMT, making it traceless. It would be interesting to see the effect of the term at two (or more) loop level.

%There is another thing that we did not discuss here but which has to be mentioned. In the case at hand both particles, corresponding to $\phi$ and $\c$ fields are massless, therefore, one should expect IR  divergences. To deal with them systematically one can introduce a small mass for each field. That would help to decouple UV and IR divergences. In the end  this mass regulator should be set to zero.

Finally, we want to stress once again that we considered the system in flat space-time.

\section{Acknowledgements}

We would like to thank R.~Rattazzi for valuable discussions. This work was supported by the Swiss National Science Foundation and by the Tomalla Foundation.

%We do not consider curved space-time neither we take into account several insertions (3-point and 4-point functions in 4 and 6 dimensions correspondingly) of the energy momentum tensor which are anomalous.

%
%{\bf It was shown that introducing a non-propagating dilaton \cite{Luty:2012ww,Komargodski:2011xv,Komargodski:2011vj,Codello:2012sn} makes the system manifestly conformal invariant even at the quantum level. 
%
%Compensators, see how the cancelation works, what modifies the current making it conserved
%
%
%Polchinski \cite{Polchinski:1987dy}
%
%
%Disclaimer regarding only scalar fields
%
%}

\appendix{Appendix: Summing all $v^{-N}$ terms}

Here we show how to get the result for the counterterms (\ref{C1_coeff}) and (\ref{C23}) for an arbitrary order in $v$. 
To renormalize the diagrams in Fig.~\ref{n6_1loop_v} of order $v^ {N}$, we consider the diagrams with $k$ and $(N-k)$ $\c$ legs in the two vertices correspondingly (see Fig.~\ref{n6}).

\begin{figure}[H]
\centering
\includegraphics[height=2.5cm]{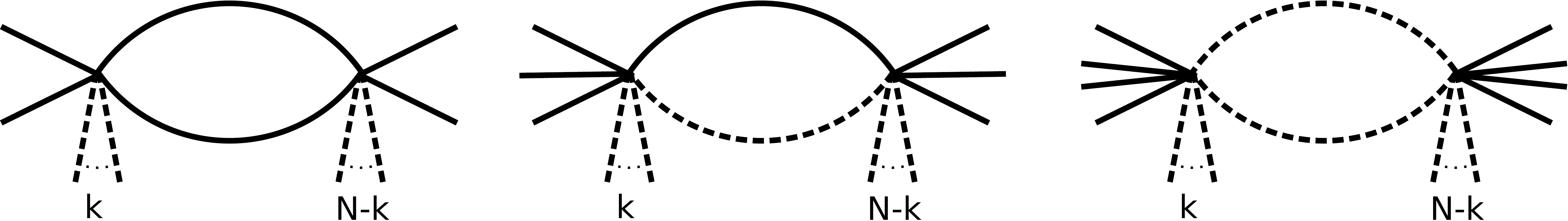}
\caption{One-loop diagrams corresponding to the correlators with $N$ $\c$ legs (dashed lines) and $4$, $6$ and $8$ $\phi$~legs (solid lines).}
\end{figure}

Using the expansion of the potential in (\ref{n6_Lagrangian})
\be
V (\phi, X)=\f{\lambda } {4!} \f{\phi ^ 4} {X } = \f{\lambda } {4!} \f{\phi ^ 4} {v } \sum _ {m=0} ^ {\infty} (-1)^m \l ( \f {\c} {v} \r ) ^ m,
\ee
one can compute the unrenormalized expression for the correlator
\be
A _ {l|N}=\la \phi _ 1 \dots \phi _ l \c _ 1 \dots \c _ N \ra.
\label{correlator}
\ee
Considering the three diagrams in Fig.~\ref{n6} with $4$, $6$ and $8$ external $\phi$ legs separately, we get
\bea
A _ {4|N} ^ 0 & = & (- 1)^ N \f {\lambda ^ 2 } {4 v^{N+2}} \sum _ {k=0} ^ N (N-k)! \, k! \sum _ {\{i\}_2\{a\}_k} G(\{i\}_2 | \{a\}_k), \nn \\
A _ {6|N} ^ 0 & = & (- 1)^ N \f {\lambda ^ 2 } {2 v^{N+4}} \sum _ {k=0} ^ N (N-k+1)! \, (k+1)! 
\sum _ {\{ i \}_3 \{ a \}_k} G(\{i\}_3 | \{a\}_k), \nn \\
A _ {8|N} ^ 0 & = & (- 1)^ N \f {\lambda ^ 2 } {4 v^{N+6}} \sum _ {k=0} ^ N (N-k+2)! \, (k+2)! 
\sum _ {\{i\}_4\{a\}_k} G(\{i\}_4 | \{a\}_k),
\eea
where $ \{\  \}_k$ is a set of $k$ numbers, and the function $G (\{i\} | \{a\}) $ is defined with the help of~(\ref{function_F}) as
\be
G (\{i\} | \{a\}) = F_ n (p_{\{i\} | \{a\}}).
\ee
On the other hand the counterterms (\ref{6n_ct}) and (\ref{6n_ct_II}) give the following contribution to the correlator~(\ref{correlator})
\bea
A _ {4|N} ^ {\text{c.t.}} & = & i (-1)^ {N+1} \f {C _ 1} {6-n} \f {\lambda ^ 2} {v ^ {N+2}} \sum _ {k=0} ^ N (N-k)! \, k! 
\sum _ {\{i\} _ 2 \{a\}_k} p ^ 2 _ {\{i\} _ 2 \{a\}_k}, \nn \\
A _ {6|N} ^ {\text{c.t.}} & = & i (-1)^ {N+1} \f {C _ 2} {6-n} \f {\lambda ^ 2} {v ^ {N+4}} \sum _ {k=0} ^ N (N-k+1)! \, (k+1)! 
\sum _ {\{i\} _ 3 \{a\}_k} p ^ 2 _ {\{i\} _ 3 \{a\}_k}, \nn \\
A _ {8|N} ^ {\text{c.t.}} & = & i (-1)^ {N+1} \f {C _ 3} {6-n} \f {\lambda ^ 2} {v ^ {N+6}} \sum _ {k=0} ^ N \f{(N-k+2)!}{2} \, \f{(k+2)!}{2}
\sum _ {\{i\} _ 4 \{a\}_k} p ^ 2 _ {\{i\} _ 4 \{a\}_k}.
\eea
Expanding the expression for $A _ {.|N} ^ 0$ around $n=4$ with the help of
\be
F _ {n} ( p ) \underset{n \to 6} \to \f {i} {3} \f {1} {(4 \pi) ^ 3} \f {p^2} {6-n},
\ee
one sees that the coefficients $C_{1,2,3}$ canceling the divergence are precisely the ones given in (\ref{C1_coeff}) and (\ref{C23}).

%\l ( p _ {12 \{a\}_k} + p _ {13 \{a\}_k} + p _ {14 \{a\}_k} + p _ {23 \{a\}_k} + p _ {24 \{a\}_k} + p _ {34 \{a\}_k} \r )

\bibliographystyle{utphys}
\bibliography{qcs_p}{}

\providecommand{\href}[2]{#2}\begingroup\raggedright\begin{thebibliography}{10}

\bibitem{Kadanoff:1966wm}
L.~Kadanoff, ``{Scaling Laws for Ising Models Near T(C)},''
{\em Physics} {\bfseries 2} (1966) 263--272.
%%CITATION = PYCSA,2,263;%%.

\bibitem{Polyakov:1970xd}
A.~M. Polyakov, ``{Conformal Symmetry of Critical Fluctuations},''
{\em JETP Lett.} {\bfseries 12} (1970) 381--383.
%%CITATION = JTPLA,12,381;%%.

\bibitem{Belavin:1984vu}
A.~Belavin, A.~M. Polyakov, and A.~Zamolodchikov, ``{Infinite Conformal
  Symmetry in Two-Dimensional Quantum Field Theory},''
\href{http://dx.doi.org/10.1016/0550-3213(84)90052-X}{{\em Nucl.Phys.}
  {\bfseries B241} (1984) 333--380}.
%%CITATION = NUPHA,B241,333;%%.

\bibitem{Komargodski:2011xv}
Z.~Komargodski, ``{The Constraints of Conformal Symmetry on RG Flows},''
  \href{http://dx.doi.org/10.1007/JHEP07(2012)069}{{\em JHEP} {\bfseries 1207}
  (2012) 069},
\href{http://arxiv.org/abs/1112.4538}{{\ttfamily arXiv:1112.4538 [hep-th]}}.
%%CITATION = ARXIV:1112.4538;%%.

\bibitem{Komargodski:2011vj}
Z.~Komargodski and A.~Schwimmer, ``{On Renormalization Group Flows in Four
  Dimensions},'' \href{http://dx.doi.org/10.1007/JHEP12(2011)099}{{\em JHEP}
  {\bfseries 1112} (2011) 099},
\href{http://arxiv.org/abs/1107.3987}{{\ttfamily arXiv:1107.3987 [hep-th]}}.
%%CITATION = ARXIV:1107.3987;%%.

\bibitem{Luty:2012ww}
M.~A. Luty, J.~Polchinski, and R.~Rattazzi, ``{The A-Theorem and the
  Asymptotics of 4D Quantum Field Theory},''
\href{http://arxiv.org/abs/1204.5221}{{\ttfamily arXiv:1204.5221 [hep-th]}}.
%%CITATION = ARXIV:1204.5221;%%.

\bibitem{Fortin:2012hc}
J.-F. Fortin, B.~Grinstein, C.~W. Murphy, and A.~Stergiou, ``{On Limit Cycles
  in Supersymmetric Theories},''
\href{http://arxiv.org/abs/1210.2718}{{\ttfamily arXiv:1210.2718 [hep-th]}}.
%%CITATION = ARXIV:1210.2718;%%.

\bibitem{Fortin:2012hn}
J.-F. Fortin, B.~Grinstein, and A.~Stergiou, ``{A Generalized C-Theorem and the
  Consistency of Scale without Conformal Invariance},''
\href{http://arxiv.org/abs/1208.3674}{{\ttfamily arXiv:1208.3674 [hep-th]}}.
%%CITATION = ARXIV:1208.3674;%%.

\bibitem{Sohnius:1981sn}
M.~F. Sohnius and P.~C. West, ``{Conformal Invariance in ${\mathcal{N}}\!=4$
  Supersymmetric Yang-Mills Theory},''
\href{http://dx.doi.org/10.1016/0370-2693(81)90326-9}{{\em Phys.Lett.}
  {\bfseries B100} (1981) 245}.
%%CITATION = PHLTA,B100,245;%%.

\bibitem{Aharony:1999ti}
O.~Aharony, S.~S. Gubser, J.~M. Maldacena, H.~Ooguri, and Y.~Oz, ``{Large $N$
  Field Theories, String Theory and Gravity},''
  \href{http://dx.doi.org/10.1016/S0370-1573(99)00083-6}{{\em Phys.Rept.}
  {\bfseries 323} (2000) 183--386},
\href{http://arxiv.org/abs/hep-th/9905111}{{\ttfamily arXiv:hep-th/9905111
  [hep-th]}}.
%%CITATION = HEP-TH/9905111;%%.

\bibitem{Wess:1960}
J.~Wess, ``{The Conformal Invariance in Quantum Field Theory},'' {\em Nuovo
  Cim.} {\bfseries 18} (1960) 1086.

\bibitem{Mack:1969rr}
G.~Mack and A.~Salam, ``{Finite Component Field Representations of the
  Conformal Group},''
\href{http://dx.doi.org/10.1016/0003-4916(69)90278-4}{{\em Annals Phys.}
  {\bfseries 53} (1969) 174--202}.
%%CITATION = APNYA,53,174;%%.

\bibitem{Ferrara:1973eg}
S.~Ferrara, R.~Gatto, and A.~Grillo, ``{Conformal Algebra in Space-Time and
  Operator Product Expansion},''
\href{http://dx.doi.org/10.1007/BFb0111104}{{\em Springer Tracts Mod.Phys.}
  {\bfseries 67} (1973) 1--64}.
%%CITATION = STPHB,67,1;%%.

\bibitem{DiFrancesco:1997nk}
P.~Di~Francesco, P.~Mathieu, and D.~Senechal,
``{Conformal Field Theory},''.
%%CITATION = INSPIRE-454643;%%.

\bibitem{Blumenhagen:2009zz}
R.~Blumenhagen and E.~Plauschinn, ``{Introduction to Conformal Field Theory},''
\href{http://dx.doi.org/10.1007/978-3-642-00450-6}{{\em Lect.Notes Phys.}
  {\bfseries 779} (2009) 1--256}.
%%CITATION = LNPHA,779,1;%%.

\bibitem{Shaposhnikov:2008xi}
M.~Shaposhnikov and D.~Zenhausern, ``{Quantum Scale Invariance, Cosmological
  Constant and Hierarchy Problem},''
  \href{http://dx.doi.org/10.1016/j.physletb.2008.11.041}{{\em Phys.Lett.}
  {\bfseries B671} (2009) 162--166},
\href{http://arxiv.org/abs/0809.3406}{{\ttfamily arXiv:0809.3406 [hep-th]}}.
%%CITATION = ARXIV:0809.3406;%%.

\bibitem{Shaposhnikov:2008xb}
M.~Shaposhnikov and D.~Zenhausern, ``{Scale Invariance, Unimodular Gravity and
  Dark Energy},'' \href{http://dx.doi.org/10.1016/j.physletb.2008.11.054}{{\em
  Phys.Lett.} {\bfseries B671} (2009) 187--192},
\href{http://arxiv.org/abs/0809.3395}{{\ttfamily arXiv:0809.3395 [hep-th]}}.
%%CITATION = ARXIV:0809.3395;%%.

\bibitem{Coleman_Aspects}
S.~R. Coleman, {\em Aspects of Symmetry}.
\newblock Cambridge University Press, 1988.

\bibitem{Coleman:1969sm}
S.~R. Coleman, J.~Wess, and B.~Zumino, ``{Structure of Phenomenological
  Lagrangians. 1.},''
\href{http://dx.doi.org/10.1103/PhysRev.177.2239}{{\em Phys.Rev.} {\bfseries
  177} (1969) 2239--2247}.
%%CITATION = PHRVA,177,2239;%%.

\bibitem{Callan:1969sn}
J.~Callan, Curtis~G., S.~R. Coleman, J.~Wess, and B.~Zumino, ``{Structure of
  Phenomenological Lagrangians. 2.},''
\href{http://dx.doi.org/10.1103/PhysRev.177.2247}{{\em Phys.Rev.} {\bfseries
  177} (1969) 2247--2250}.
%%CITATION = PHRVA,177,2247;%%.

\bibitem{Salam:1969rq}
A.~Salam and J.~Strathdee, ``{Nonlinear Realizations. 1: the Role of Goldstone
  Bosons},''
\href{http://dx.doi.org/10.1103/PhysRev.184.1750}{{\em Phys.Rev.} {\bfseries
  184} (1969) 1750--1759}.
%%CITATION = PHRVA,184,1750;%%.

\bibitem{Salam:1970qk}
A.~Salam and J.~Strathdee, ``{Nonlinear Realizations. 2. Conformal Symmetry},''
\href{http://dx.doi.org/10.1103/PhysRev.184.1760}{{\em Phys.Rev.} {\bfseries
  184} (1969) 1760--1768}.
%%CITATION = PHRVA,184,1760;%%.

\bibitem{Low:2001bw}
I.~Low and A.~V. Manohar, ``{Spontaneously Broken Space-Time Symmetries and
  Goldstone's Theorem},''
  \href{http://dx.doi.org/10.1103/PhysRevLett.88.101602}{{\em Phys.Rev.Lett.}
  {\bfseries 88} (2002) 101602},
\href{http://arxiv.org/abs/hep-th/0110285}{{\ttfamily arXiv:hep-th/0110285
  [hep-th]}}.
%%CITATION = HEP-TH/0110285;%%.

\bibitem{Isham:1970gz}
C.~Isham, A.~Salam, and J.~Strathdee, ``{Spontaneous Breakdown of Conformal
  Symmetry},''
\href{http://dx.doi.org/10.1016/0370-2693(70)90177-2}{{\em Phys.Lett.}
  {\bfseries B31} (1970) 300--302}.
%%CITATION = PHLTA,B31,300;%%.

\bibitem{Rattazzi:2003ea}
R.~Rattazzi, ``{Cargese Lectures on Extra-Dimensions},''
\href{http://arxiv.org/abs/hep-ph/0607055}{{\ttfamily arXiv:hep-ph/0607055
  [hep-ph]}}.
%%CITATION = HEP-PH/0607055;%%.

\bibitem{Sundrum:1998sj}
R.~Sundrum, ``{Effective Field Theory for a Three-Brane Universe},''
  \href{http://dx.doi.org/10.1103/PhysRevD.59.085009}{{\em Phys.Rev.}
  {\bfseries D59} (1999) 085009},
\href{http://arxiv.org/abs/hep-ph/9805471}{{\ttfamily arXiv:hep-ph/9805471
  [hep-ph]}}.
%%CITATION = HEP-PH/9805471;%%.

\bibitem{Englert:1976ep}
F.~Englert, C.~Truffin, and R.~Gastmans, ``{Conformal Invariance in Quantum
  Gravity},''
\href{http://dx.doi.org/10.1016/0550-3213(76)90406-5}{{\em Nucl.Phys.}
  {\bfseries B117} (1976) 407}.
%%CITATION = NUPHA,B117,407;%%.

\bibitem{Wetterich:1987fm}
C.~Wetterich, ``{Cosmology and the Fate of Dilatation Symmetry},''
\href{http://dx.doi.org/10.1016/0550-3213(88)90193-9}{{\em Nucl.Phys.}
  {\bfseries B302} (1988) 668}.
%%CITATION = NUPHA,B302,668;%%.

\bibitem{Wetterich:1987fk}
C.~Wetterich, ``{Cosmologies with Variable Newton's `Constant'},''
\href{http://dx.doi.org/10.1016/0550-3213(88)90192-7}{{\em Nucl.Phys.}
  {\bfseries B302} (1988) 645}.
%%CITATION = NUPHA,B302,645;%%.

\bibitem{Capper:1974ic}
D.~Capper and M.~Duff, ``{Trace Anomalies in Dimensional Regularization},''
\href{http://dx.doi.org/10.1007/BF02748300}{{\em Nuovo Cim.} {\bfseries A23}
  (1974) 173--183}.
%%CITATION = NUCIA,A23,173;%%.

\bibitem{Duff:1993wm}
M.~Duff, ``{Twenty Years of the Weyl Anomaly},''
  \href{http://dx.doi.org/10.1088/0264-9381/11/6/004}{{\em Class.Quant.Grav.}
  {\bfseries 11} (1994) 1387--1404},
\href{http://arxiv.org/abs/hep-th/9308075}{{\ttfamily arXiv:hep-th/9308075
  [hep-th]}}.
%%CITATION = HEP-TH/9308075;%%.

\bibitem{Shaposhnikov:2008ar}
M.~E. Shaposhnikov and I.~I. Tkachev, ``{Quantum Scale Invariance on the
  Lattice},'' \href{http://dx.doi.org/10.1016/j.physletb.2009.04.040}{{\em
  Phys.Lett.} {\bfseries B675} (2009) 403--406},
\href{http://arxiv.org/abs/0811.1967}{{\ttfamily arXiv:0811.1967 [hep-th]}}.
%%CITATION = ARXIV:0811.1967;%%.

\bibitem{Polchinski:1987dy}
J.~Polchinski, ``{Scale and Conformal Invariance in Quantum Field Theory},''
\href{http://dx.doi.org/10.1016/0550-3213(88)90179-4}{{\em Nucl.Phys.}
  {\bfseries B303} (1988) 226}.
%%CITATION = NUPHA,B303,226;%%.

\bibitem{Shaposhnikov:2009nk}
M.~Shaposhnikov and F.~Tkachov, ``{Quantum Scale-Invariant Models as Effective
  Field Theories},''
\href{http://arxiv.org/abs/0905.4857}{{\ttfamily arXiv:0905.4857 [hep-th]}}.
%%CITATION = ARXIV:0905.4857;%%.

\bibitem{Brown:1979pq}
L.~S. Brown, ``{Dimensional Regularization of Composite Operators in Scalar
  Field Theory},''
\href{http://dx.doi.org/10.1016/0003-4916(80)90377-2}{{\em Annals Phys.}
  {\bfseries 126} (1980) 135}.
%%CITATION = APNYA,126,135;%%.

\bibitem{Collins_book}
J.~Collins, {\em Renormalization: An Introduction to Renormalization, the
  Renormalization Group and the Operator-Product Expansion}.
\newblock Cambridge University Press, 1984.

\bibitem{Codello:2012sn}
A.~Codello, G.~D'Odorico, C.~Pagani, and R.~Percacci, ``{The Renormalization
  Group and Weyl-Invariance},''
\href{http://arxiv.org/abs/1210.3284}{{\ttfamily arXiv:1210.3284 [hep-th]}}.
%%CITATION = ARXIV:1210.3284;%%.

\bibitem{Adler:1976zt}
S.~L. Adler, J.~C. Collins, and A.~Duncan, ``{Energy-Momentum-Tensor Trace
  Anomaly in Spin 1/2 Quantum Electrodynamics},''
\href{http://dx.doi.org/10.1103/PhysRevD.15.1712}{{\em Phys.Rev.} {\bfseries
  D15} (1977) 1712}.
%%CITATION = PHRVA,D15,1712;%%.

\bibitem{Giannotti:2008cv}
M.~Giannotti and E.~Mottola, ``{The Trace Anomaly and Massless Scalar Degrees
  of Freedom in Gravity},''
  \href{http://dx.doi.org/10.1103/PhysRevD.79.045014}{{\em Phys.Rev.}
  {\bfseries D79} (2009) 045014},
\href{http://arxiv.org/abs/0812.0351}{{\ttfamily arXiv:0812.0351 [hep-th]}}.
%%CITATION = ARXIV:0812.0351;%%.

\end{thebibliography}\endgroup

\end{document}